\documentclass[12pt]{article}
\usepackage{a4wide,epsfig}

\usepackage{slashed}

\newcommand{\agt}{\rlap{\lower 3.5 pt \hbox{$\mathchar \sim$}} \raise 1pt
 \hbox {$>$}}
\newcommand{\alt}{\rlap{\lower 3.5 pt \hbox{$\mathchar \sim$}} \raise 1pt
 \hbox {$<$}}

\catcode`@=11
\newcount\@tempcntc
\def\@citex[#1]#2{\if@filesw\immediate\write\@auxout{\string\citation{#2}}\fi
  \@tempcnta\z@\@tempcntb\m@ne\def\@citea{}\@cite{\@for\@citeb:=#2\do
    {\@ifundefined
       {b@\@citeb}{\@citeo\@tempcntb\m@ne\@citea\def\@citea{,}{\bf
?}\@warning
       {Citation `\@citeb' on page \thepage \space undefined}}%
    {\setbox\z@\hbox{\global\@tempcntc0\csname b@\@citeb\endcsname\relax}%
     \ifnum\@tempcntc=\z@ \@citeo\@tempcntb\m@ne
       \@citea\def\@citea{,}\hbox{\csname b@\@citeb\endcsname}%
     \else
      \advance\@tempcntb\@ne
      \ifnum\@tempcntb=\@tempcntc
      \else\advance\@tempcntb\m@ne\@citeo
      \@tempcnta\@tempcntc\@tempcntb\@tempcntc\fi\fi}}\@citeo}{#1}}
\def\@citeo{\ifnum\@tempcnta>\@tempcntb\else\@citea\def\@citea{,}%
  \ifnum\@tempcnta=\@tempcntb\the\@tempcnta\else
   {\advance\@tempcnta\@ne\ifnum\@tempcnta=\@tempcntb \else
\def\@citea{--}\fi
    \advance\@tempcnta\m@ne\the\@tempcnta\@citea\the\@tempcntb}\fi\fi}
\catcode`@=12

\begin{document}

\title{
\vskip-3cm{\baselineskip14pt
\centerline{\normalsize DESY 11--157\hfill ISSN 0418-9833}
\centerline{\normalsize September 2011\hfill}}
\vskip1.5cm
Probing nonrelativistic QCD factorization in polarized $J/\psi$ photoproduction
at next-to-leading order}

\author{Mathias Butenschoen, Bernd A. Kniehl\\
{\normalsize II. Institut f\"ur Theoretische Physik, Universit\"at Hamburg,}\\
{\normalsize Luruper Chaussee 149, 22761 Hamburg, Germany}
}
\date{}

\maketitle

\begin{abstract}
We analyze the polarization observables of $J/\psi$ photoproduction at
next-to-leading order (NLO) within the factorization formalism of
nonrelativistic quantum chromodynamics (NRQCD).
This is the first NLO study of heavy-quarkonium polarization including the full
relativistic corrections due to the intermediate $^1\!S_0^{[8]}$,
$^3\!S_1^{[8]}$, and $^3\!P_J^{[8]}$ color-octet (CO) states in the worldwide
endeavor to test NRQCD factorization at the quantum level.
We present theoretical predictions in the helicity, target, and Collins-Soper
frames of DESY HERA, evaluated using the CO long-distance matrix elements
previously extracted through a global fit to experimental data of unpolarized
$J/\psi$ production, and confront them with recent measurements by the H1 and
ZEUS Collaborations.
We find the overall agreement to be satisfactory, but the case for NRQCD to be
not as strong as for the $J/\psi$ yield.

\medskip

\end{abstract}

\newpage

The test of NRQCD factorization \cite{Bodwin:1994jh} has been identified to be
among the most exigent milestones on the roadmap of heavy-quarkonium physics at
the present time \cite{Brambilla:2010cs}.
Quarkonia are systems consisting of a quark and its antiparticle bound by the
strong force, among which charmonium $c\overline{c}$ and bottomonium
$b\overline{b}$ are considered heavy.
The $J/\psi$ meson, the lowest-lying $c\overline{c}$ state of spin one, which
was simultaneously discovered at the Brookhaven National Laboratory
\cite{Aubert:1974js} and the Stanford Linear Accelerator Center
\cite{Augustin:1974xw} in November 1974 (The Nobel Prize in Physics 1976),
provides a particularly useful laboratory for such a test because it is
copiously produced at all high-energy particle colliders, owing to its
relatively low mass, and particularly easy to detect experimentally. 
In fact, sharing the total-angular-momentum, parity, and charge-conjugation
quantum numbers $J^{PC}=1^{--}$ with the photon, it can decay to $e^+e^-$ and
$\mu^+\mu^-$ pairs producing spectacular signatures in the detectors, the
branching fraction of either decay channel being as large as about 6\%
\cite{Nakamura:2010zzi}.

In fact, the NRQCD factorization formalism \cite{Bodwin:1994jh} provides a
rigorous theoretical framework for the description of heavy-quarkonium
production and decay.
This implies a separation of process-dependent short-distance coefficients, to
be calculated perturbatively as expansions in the strong-coupling constant
$\alpha_s$, from supposedly universal long-distance matrix elements
(LDMEs), to be extracted from experiment.
The relative importance of the latter can be estimated by means of velocity
scaling rules, which predict each of the LDMEs to scale with a definite
power of the heavy-quark ($Q=c,b$) velocity $v$ in the limit $v\ll1$.
In this way, the theoretical predictions are organized as double expansions in
$\alpha_s$ and $v$.
A crucial feature of this formalism is that the $Q\overline{Q}$ pair can at
short distances be produced in any Fock state
$n={}^{2S+1}L_J^{[a]}$ with definite spin $S$, orbital angular momentum
$L$, total angular momentum $J$, and color multiplicity $a=1,8$.
In particular, this formalism predicts the existence of intermediate CO states
in nature, which subsequently evolve into physical,
color-singlet (CS) quarkonia by the nonperturbative emission of soft gluons.
In the limit $v\to0$, the traditional CS model (CSM) is recovered in the case
of $S$-wave quarkonia.
In the case of $J/\psi$ production, the CSM prediction is based just on the
$^3\!S_1^{[1]}$ CS state, while the leading relativistic corrections, of
relative order ${\cal O}(v^4)$, are built up by the $^1\!S_0^{[8]}$,
$^3\!S_1^{[8]}$, and $^3\!P_J^{[8]}$ ($J=0,1,2$) CO states.

The CSM is not a complete theory, as may be understood by noticing that the NLO
treatment of $P$-wave quarkonia is plagued by uncanceled infrared
singularities, which are, however, properly removed in NRQCD.
This conceptual problem cannot be cured from within the CSM, neither by
proceeding to higher orders nor by invoking $k_T$ factorization {\it etc.}.
As it were, NRQCD factorization, appropriately improved at large transverse
momenta $p_T$ by systematic expansion in powers of $m_Q^2/p_T^2$
\cite{Kang:2011zz}, is the only game in town, which makes its
experimental verification such a matter of paramount importance and general
interest \cite{Brambilla:2010cs}.

The present status of testing NRQCD factorization in charmonium production is
as follows.
Very recently, NRQCD factorization has been consolidated at NLO by a global fit
\cite{Butenschoen:2011yh} to all available high-quality data of inclusive
unpolarized $J/\psi$ production, comprising a total of 194 data points from 26
data sets collected by 10 experiments at 6 colliders, namely by Belle at KEKB;
DELPHI at LEP~II; H1 and ZEUS at HERA~I and II; PHENIX at RHIC; CDF at
Tevatron~I and II; and ATLAS, CMS, ALICE, and LHCb at the LHC.
This fit successfully pinned down the three CO LDMEs in compliance with the
velocity scaling rules, establishing their universality, and yielded an overall
description of the data well within the theoretical uncertainties; appreciable
deviations arose only in the case of two-photon scattering, where the useable
data comprises only 16 events and has not been confirmed by any of the other
three LEP~II experiments, however.
On the other hand, the NLO CS predictions were found to significantly
undershoot all the measurements, except for the single data point of $e^+e^-$
annihilation.

In contrast to the $J/\psi$ yield, NRQCD interpretations of $J/\psi$
polarization measurements have so far been exhibiting a rather confusing
pattern \cite{Brambilla:2010cs}, presumably because the theoretical status is
much less advanced there.
In fact, complete NRQCD predictions for $J/\psi$ polarization observables so
far only exist at LO.
At NLO, the CSM predictions for direct photoproduction
\cite{Artoisenet:2009xh,Chang:2009uj} and hadroproduction \cite{Gong:2008sn} as
well as the $^1\!S_0^{[8]}$ and $^3\!S_1^{[8]}$ contributions to
hadroproduction \cite{Gong:2008ft}, which may be obtained using standard
techniques, are known.
The NLO calculation of $^3\!P_J^{[8]}$ contributions, which are expected to be
significant, is far more intricate because the applications of the respective
projection operators to the short-distance scattering amplitudes produce
particularly lengthy expressions involving complicated tensor loop integrals
and exhibiting an entangled pattern of infrared singularities.
This technical bottleneck is overcome here for the first time for $J/\psi$
polarization observables.

Recent high-quality measurements by the H1 \cite{Aaron:2010gz} and ZEUS
\cite{:2009br} Collaborations at HERA provide a strong motivation for us to
start by studying photoproduction, where the incoming leptons interact with the
protons via quasi-real photons, of low virtuality $Q^2=-p_\gamma^2$, and are
deflected under small angles.
Such quasi-real photons participate in the hard scattering either directly or
via partons into which they fluctuate (resolve) intermittently.
However, resolved photoproduction is greatly suppressed, to the level of 1\%
\cite{Butenschoen:2011yh}, by the cut $z>0.3$ (0.4) applied by H1
\cite{Aaron:2010gz} (ZEUS \cite{:2009br}) and is thus neglected here.
Here, $z=(p_{J/\psi}\cdot p_p)/(p_{\gamma}\cdot p_p)$, with $p_{J/\psi}$,
$p_{\gamma}$, and $p_p$ being the four-momenta of the $J/\psi$ meson, photon,
and proton, respectively, denotes the inelasticity variable, which measures the
fraction of photon energy transferred to the $J/\psi$ meson in the proton rest
frame.
Another important variable of photoproduction is the $\gamma p$ invariant mass,
$W=\sqrt{(p_\gamma+p_p)^2}$.

The polarization of the $J/\psi$ meson is conveniently analyzed experimentally
by measuring the angular distribution of its leptonic decays, which is
customarily parametrized using the three polarization observables
$\lambda$, $\mu$, and $\nu$, as \cite{Lam:1978pu}
\begin{eqnarray}
\frac{d\Gamma(J/\psi\to l^+l^-)}{d\cos\theta\,d\phi}&\propto&
1 + \lambda \cos^2 \theta + \mu \sin(2\theta) \cos \phi \nonumber \\
 &&{} + \frac{\nu}{2} \sin^2\theta \cos (2\phi),
\end{eqnarray}
where $\theta$ and $\phi$ are respectively the polar the azimuthal angles of
$l^+$ in the $J/\psi$ rest frame.
This definition, of course, depends on the choice of coordinate frame.
Among the most frequently employed coordinate frames are the helicity (recoil),
Collins-Soper, and target frames, in which the polar axes point in the
direction of $-(\vec{p}_\gamma+\vec{p}_p)$,
$\vec{p}_\gamma/|\vec{p}_\gamma|-\vec{p}_p/|\vec{p}_p|$, and $-\vec{p}_p$,
respectively.
The values $\lambda=0,+1,-1$ correspond to unpolarized, fully transversely
polarized, and fully longitudinally polarized $J/\psi$ mesons, respectively.

On the theoretical side, we have
\begin{eqnarray}
 \lambda&=& \frac{d\sigma_{11} - d\sigma_{00}}{d\sigma_{11}+d\sigma_{00}},
\nonumber \\
 \mu &=& \frac{\sqrt{2}\mbox{Re}\,d\sigma_{10}}{d\sigma_{11}+d\sigma_{00}},
\nonumber \\
 \nu &=& \frac{2 d\sigma_{1,-1}}{d\sigma_{11}+d\sigma_{00}},
\label{eq:lmn}
\end{eqnarray}
where $d\sigma_{ij}$, with $i,j=0,\pm1$ denoting the $z$ component of $S$, is
the $ij$ component of the $ep\to J/\psi+X$ differential cross section in the
spin density matrix formalism.
Invoking the Weizs\"acker-Williams approximation and the factorization theorems
of the QCD parton model and NRQCD \cite{Bodwin:1994jh}, we have
\begin{eqnarray}
 d\sigma_{ij} &=& \sum_{k,n} \int dxdy\, f_{\gamma/e}(x) f_{k/p}(y)
\langle{\cal O}^{J/\psi}[n]\rangle
\nonumber\\
 &&{}\times \frac{1}{2s} d\mbox{PS}\,
\overline{\rho_{ij}(\gamma k\to c\overline{c}[n]+X)},
\end{eqnarray}
where $f_{\gamma/e}(x)$ is the photon flux function, $f_{k/p}(y)$ the parton
distribution function (PDF) of parton $k=g,q,\overline{q}$ with 
$q=u,d,s$, $\langle{\cal O}^{J/\psi}[n]\rangle$ are the LDMEs,
$s=(p_\gamma+p_k)^2$, and $d\mbox{PS}$ is the phase space measure of the 
outgoing particles.
The spin density matrix elements of the partonic cross sections,
$\rho_{ij}(\gamma k\to c\overline{c}[n]+X)$, are averaged (summed) over the
spins and colors of the incoming (outgoing) particles, keeping $i$ and $j$
fixed for the $c\overline{c}$ pair in Fock state $n$.
The quantities $\rho_{ij}$ are evaluated by applying polarization and color
projectors similar to those listed in Ref.~\cite{Petrelli:1997ge} to the
squared QCD matrix elements of open $c\overline{c}$ production.
For $n={}^3\!S_1^{[1]},{}^3\!S_1^{[8]},{}^3\!P_J^{[8]}$, the $c\overline{c}[n]$
spin polarization vectors $\epsilon(i)$ appearing in $\rho_{ij}$ are replaced
by their explicit expressions \cite{Beneke:1998re}.
In the case of $n={}^3\!P_J^{[8]}$, for which $S=L=1$, the $z$ components of
$L$ are summed over.
For $n={}^1\!S_0^{[8]}$, $\rho_{11}$ and $\rho_{00}$ are each set to one third
of the squared matrix element, and $\rho_{10}=\rho_{1,-1}=0$.
For space limitation, we refrain from presenting here more technical details,
but refer the interested reader to a forthcoming publication.

\begin{table}
\begin{center}
\begin{tabular}{|c|c|}
\hline
$\langle {\cal O}^{J/\psi}(^1\!S_0^{[8]}) \rangle$ &
$(3.04\pm0.35)\times10^{-2}$~GeV$^3$ \\
$\langle {\cal O}^{J/\psi}(^3\!S_1^{[8]}) \rangle$ &
$(1.68\pm0.46)\times10^{-3}$~GeV$^3$ \\
$\langle {\cal O}^{J/\psi}(^3\!P_0^{[8]}) \rangle$ &
$(-9.08\pm1.61)\times10^{-3}$~GeV$^5$ \\
\hline
\end{tabular}
\end{center}
\caption{\label{tab:fit}%
$J/\psi$ NLO CO LDMEs corrected for feed-down \cite{Butenschoen:2011yh}.}
\end{table}

\begin{figure}[ht!]
\begin{tabular}{|c|c|c|}
\hline
\includegraphics[width=0.305\textwidth]{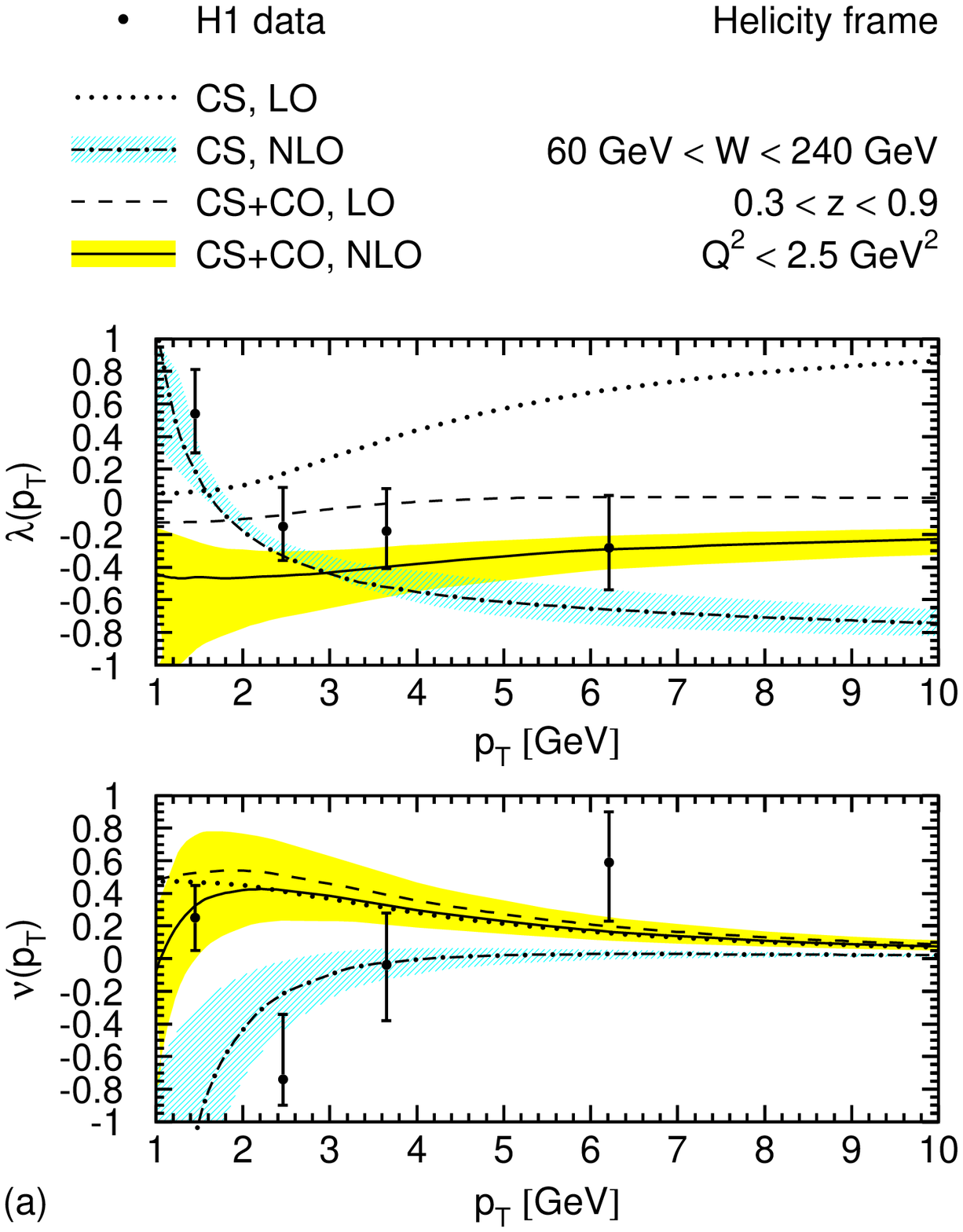}
&
\includegraphics[width=0.305\textwidth]{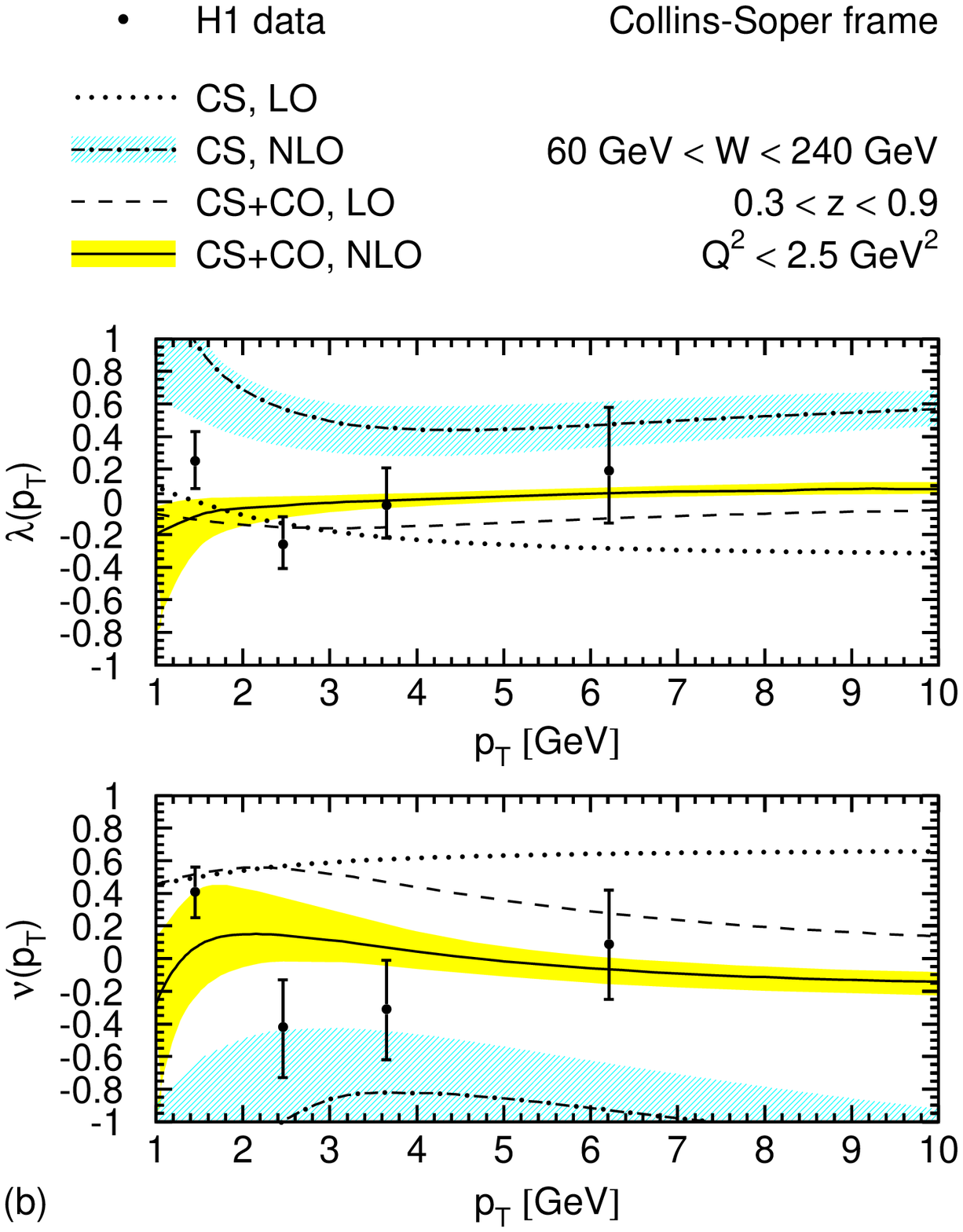}
&
\includegraphics[width=0.305\textwidth]{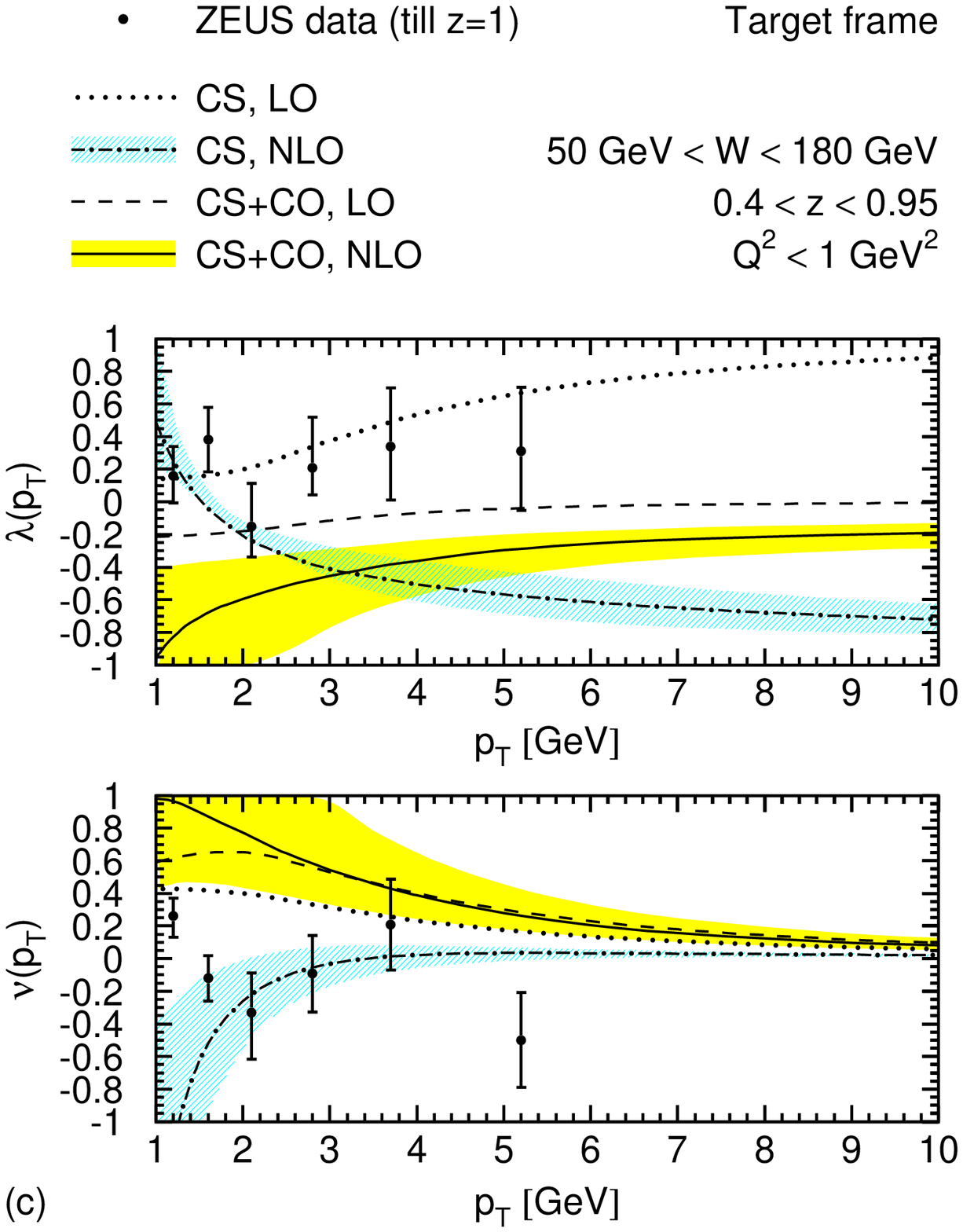}
\\ \hline
\includegraphics[width=0.305\textwidth]{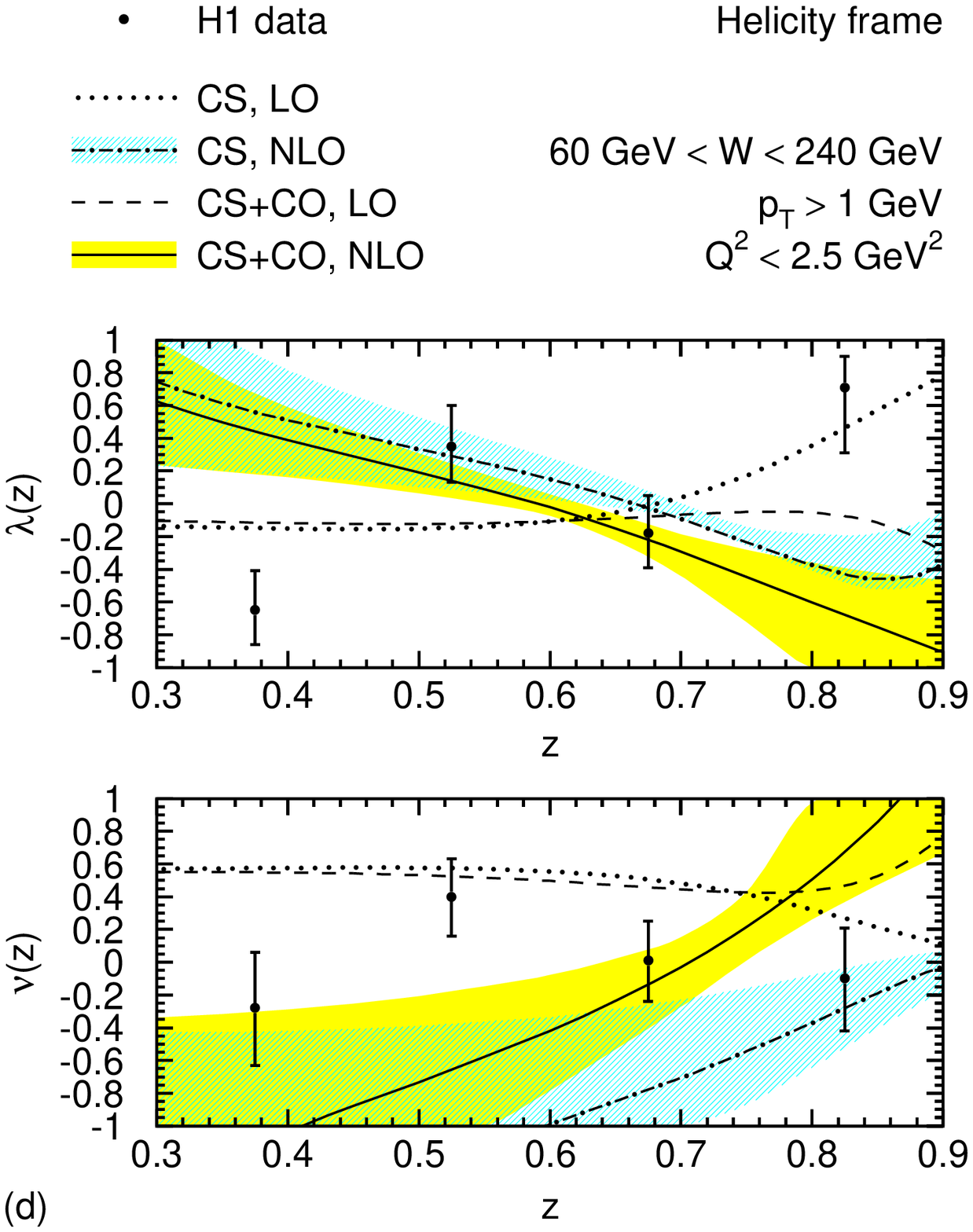}
&
\includegraphics[width=0.305\textwidth]{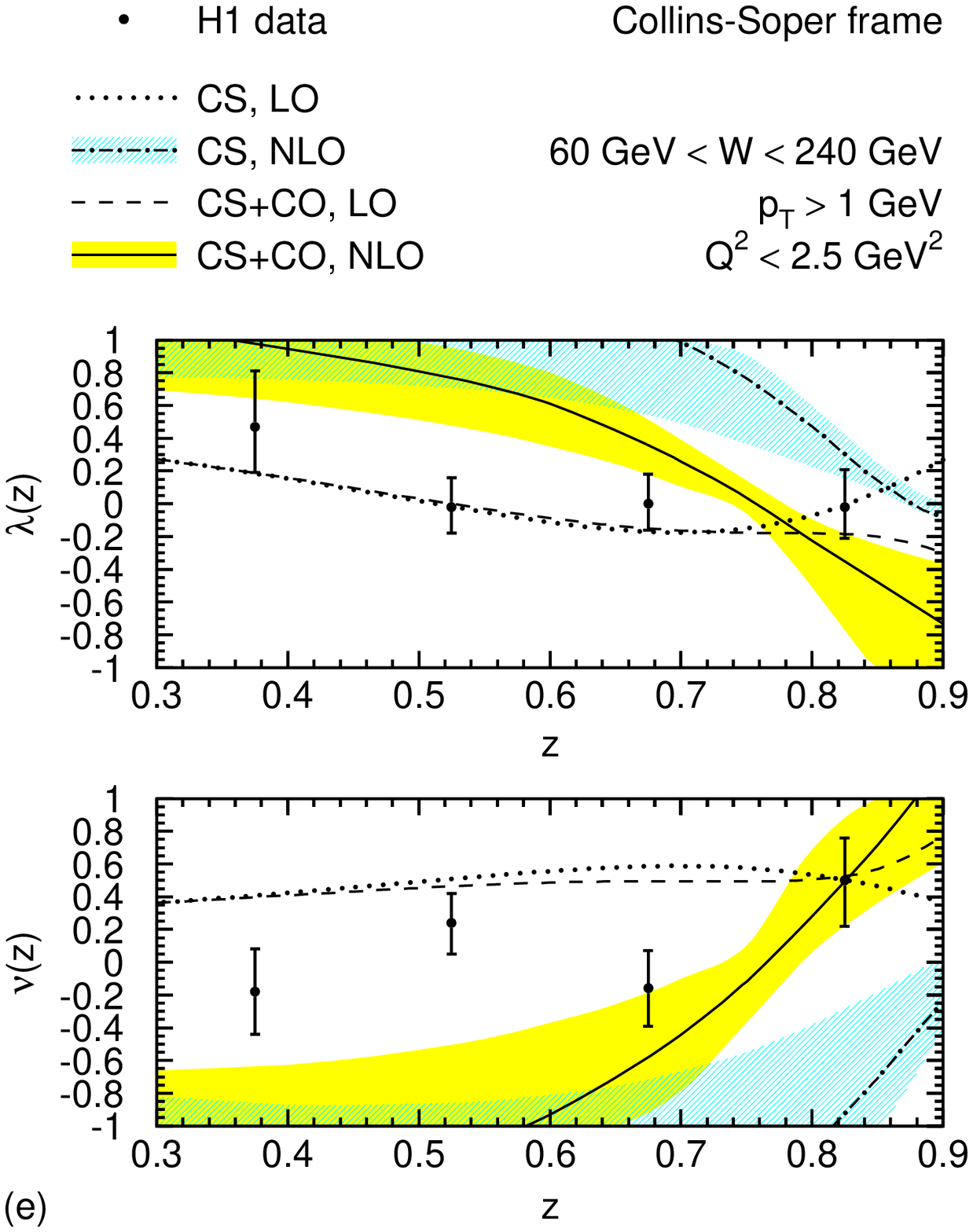}
&
\includegraphics[width=0.305\textwidth]{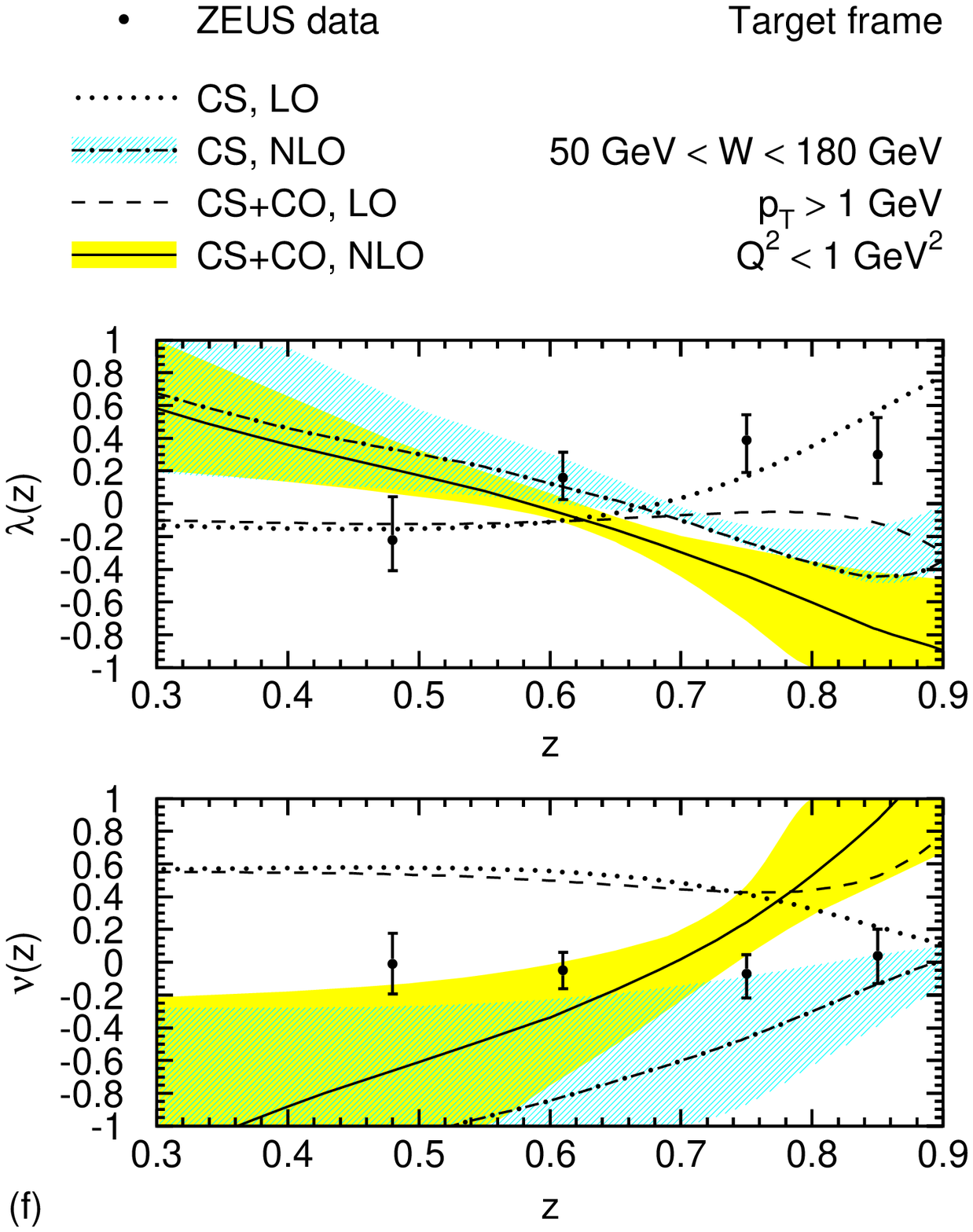}
\\ \hline
\end{tabular}
\caption{\label{fig:thegraphs}%
(color online) NLO NRQCD predictions (solid lines) for $\lambda$ and $\nu$ as
functions of $p_T$ and $z$ in the helicity, Collins-Soper, and target frames
including theoretical uncertainties (shaded/yellow bands) compared to H1
\cite{Aaron:2010gz} and ZEUS \cite{:2009br} data.
For comparison, also the NLO CSM (dot-dashed lines) predictions including
theoretical uncertainties (hatched/blue bands) as well as the LO NRQCD (dashed
lines) and LO CSM (dotted lines) ones are shown.
}
\end{figure}

We now describe the theoretical input for our numerical analysis.
In all our NRQCD calculations, we use the CO LDME set extracted in
Ref.~\cite{Butenschoen:2011yh} after subtracting from the data fitted to the
estimated contributions due to feed-down from heavier charmonia.
For the reader's convenience, these values are listed in Table~\ref{tab:fit}.
For consistency, we also adopt the residual input from
Ref.~\cite{Butenschoen:2011yh}.
In particular, we choose the CS LDME to be
$\langle {\cal O}^{J/\psi}(^3\!S_1^{[1]}) \rangle = 1.32$~GeV$^3$ 
\cite{Bodwin:2007fz} and the charm-quark mass, which we renormalize according
to the on-shell scheme, to be $m_c=1.5$~GeV, adopt the values of the electron
mass $m_e$ and the electromagnetic coupling constant $\alpha$ from
Ref.~\cite{Nakamura:2010zzi}, and use the one-loop (two-loop) formula for
$\alpha_s^{(n_f)}(\mu_r)$, with $n_f=4$ active quark
flavors, at LO (NLO).
As for the proton PDFs, we use the CTEQ6L1 (CTEQ6M) set \cite{Pumplin:2002vw}
at LO (NLO), which comes with an asymptotic scale parameter of
$\Lambda_\mathrm{QCD}^{(4)}=215$~MeV (326~MeV).
We evaluate the photon flux function using Eq.~(5) of
Ref.~\cite{Kniehl:1996we}.
Our default choices for the $\overline{\mbox{MS}}$ renormalization,
factorization, and NRQCD scales are $\mu_r=\mu_f=m_T$ and $\mu_\Lambda=m_c$,
respectively, where $m_T=\sqrt{p_T^2+4m_c^2}$ is the $J/\psi$ transverse mass.
The bulk of the theoretical uncertainty is due to the lack of knowledge of
corrections beyond NLO, which are estimated by varying $\mu_r$, $\mu_f$, and
$\mu_\Lambda$ by a factor 2 up and down relative to their default values.
In our NLO NRQCD predictions, we must also include the errors in the CO LDMEs.
To this end, we determine the maximum upward and downward shifts generated by
independently varying their values according to Table~\ref{tab:fit} and add
the resulting half-errors in quadrature to those due to scale variations.

In Fig.~\ref{fig:thegraphs}, we compare our NLO NRQCD predictions for $\lambda$
and $\nu$ as functions of $p_T$ and $z$, evaluated from Eq.~(\ref{eq:lmn}) with
the respective differential cross sections inserted on the r.h.s., with the
measurements by H1 \cite{Aaron:2010gz} in the helicity and Collins-Soper frames
and by ZEUS \cite{:2009br} in the target frame.
The H1 data was taken during the years 2006 and 2007, and corresponds to an
integrated luminosity of 165~pb$^{-1}$, while the ZEUS analysis covers all
data collected from 1996 through 2007, corresponding to 430~pb$^{-1}$.
At HERA, 27.5~GeV electrons or positrons were colliding with 820~GeV (920~GeV)
protons before (since) 1998.
As the admixture of 820~GeV protons in the ZEUS data sample is negligible, we
take the c.m.\ energy to be 318~GeV also there.
We adopt the experimental acceptance cuts, indicated in each of the six frames
of Fig.~\ref{fig:thegraphs}, except for the $p_T$ distribution by ZEUS in
Fig.~\ref{fig:thegraphs}(c).
Unfortunately, ZEUS did not impose any upper $z$ cut, which poses two problems
on the theoretical side.
On the one hand, in the kinematic endpoint region, at $z\approx1$, where the
scattering becomes elastic, the cross section is overwhelmed by diffractive
$J/\psi$ production, the treatment of which lies beyond the scope of our paper.
On the other hand, the NRQCD expansion in $v$ breaks down in the limit
$z\to1$, so that our fixed-order calculation becomes invalid.
We avoid these problems by introducing the cut $z<0.95$, accepting that the
comparison with the ZEUS data then has to be taken with a grain of salt.

For comparison, also the LO NRQCD as well as the LO and NLO CSM predictions
are shown in Fig.~\ref{fig:thegraphs}.
In order to visualize the size of the NLO corrections to the hard-scattering
cross sections, the LO predictions are evaluated with the same LDMEs.
We observe that, in all the cases considered, the inclusion of the NLO
corrections has a considerably less dramatic effect in NRQCD than in the CSM,
where the normalizations and shapes of the various distributions are radically
modified.
This indicates that the perturbative expansion in $\alpha_s$ converges more
rapidly in NRQCD than in the CSM.
Looking at the $\lambda(p_T)$ distributions in
Figs.~\ref{fig:thegraphs}(a)--(c), we notice that NRQCD predicts large-$p_T$
$J/\psi$ mesons to be approximately unpolarized, both at LO and NLO, which is
nicely confirmed by the H1 measurements in Figs.~\ref{fig:thegraphs}(a) and
(b).
However, the ZEUS measurement in Fig.~\ref{fig:thegraphs}(c), which reaches all
the way up to $z=1$, exhibits a conspicuous tendency towards transverse
polarization, which might well reflect the notion that diffractively produced
vector mesons prefer to be strongly transversely polarized at $z\approx1$
\cite{Brodsky:1994kf}.
Comparing the NLO NRQCD and CSM predictions in the three different frames, we
conclude that the Colins-Soper frame possesses the most discriminating power.
As expected, the theoretical uncertainties, which are chiefly due to scale
variations, steadily decrease as the value of $p_T$ increases, which just
reflects asymptotic freedom.
By the same token, the theoretical uncertainties in the $z$ distributions in
Figs.~\ref{fig:thegraphs}(d)--(e), which are dominated by contributions from
the $p_T$ region close to the lower cut-off at $p_T=1$~GeV, are quite sizable,
which makes a useful interpretation of the experimental data more difficult.

At this point, we compare our results with the theoretical literature.
We agree with the LO NRQCD formulas for
$\overline{\rho_{ij}(\gamma k\to c\overline{c}[n]+k)}$ listed in Appendix B of
Ref.~\cite{Beneke:1998re}.
We are able to nicely reproduce the NLO CSM results for $\lambda$ and $\nu$ as
functions of $p_T$ and $z$ shown in Fig.~2 of Ref.~\cite{Artoisenet:2009xh}
and Fig.~2 of Ref.~\cite{Chang:2009uj} if we adopt the theoretical inputs
specified there.
The differences between those NLO CSM results and the respective results in our
Fig.~\ref{fig:thegraphs} are due to the use of different theoretical inputs.
A similar statement applies to the LO NRQCD results graphically displayed in
Refs.~\cite{Aaron:2010gz,:2009br,Beneke:1998re}, which are evaluated using
CO LDMEs obtained from LO fits to Tevatron~I data.

In contrast to the unpolarized $J/\psi$ yield, where the most precise world
data uniformly and vigorously support NRQCD and distinctly disfavor the CSM at
NLO \cite{Butenschoen:2011yh}, the situation seems to be less obvious for the
$J/\psi$ polarization in photoproduction, as a superficial glance at
Fig.~\ref{fig:thegraphs} suggests.
However, detailed investigation reveals that the overall $\chi^2$ value of all
the H1 and ZEUS data in Fig.~\ref{fig:thegraphs} w.r.t.\ the default NLO
predictions is reduced by more than 50\% as the CO contributions are included,
marking a general trend towards continued verification of NRQCD factorization.
Unfortunately, this is where the legacy of HERA, which was shut down in 2007,
ends.
With the help of the proposed lepton-proton collider LHeC at CERN, polarized
$J/\psi$ photoproduction could be studied more precisely and up to much larger
values of $p_T$.
Fortunately, measurements of $J/\psi$ polarization have also been performed in
hadroproduction at the Tevatron and will be carried on at the LHC for many
years.
This is arguably the last frontier in the international endeavor to test NRQCD
factorization in charmonium physics.

This work was supported in part by BMBF Grant No.\ 05H09GUE and HGF Grant No.\
HA~101.

\end{document}